\documentstyle[twoside,fleqn,espcrc2,epsfig]{article}

\makeatletter
\newbox\slashbox \setbox\slashbox=\hbox{\large$/$}
\def\pslash#1{\setbox\@tempboxa=\hbox{$#1$}
  \@tempdima=0.5\wd\slashbox \advance\@tempdima 0.5\wd\@tempboxa
  \copy\slashbox \kern-\@tempdima \box\@tempboxa}

\title{Dirac and Gor'kov spectra in two color QCD with chemical potential}

\author{Elmar Bittner\address{Institut f\"ur Kernphysik, Technische Universit\"at Wien,
        A-1040 Vienna, Austria}, Maria-Paola Lombardo\address{Istituto Nazionale di Fisica
        Nucleare, Sezione di Padova, e Gruppo Collegato di Trento, Italy},
        Harald Markum$^{\rm a}$, and Rainer Pullirsch$^{\rm a}$}
\begin{document}
\begin{abstract}
We analyze the eigenvalue spectrum of the staggered Dirac matrix 
in two-color QCD at nonzero baryon density when the 
eigenvalues become complex. The quasi-zero modes and their role for 
chiral symmetry breaking and the deconfinement transition are examined. 
The bulk of the spectrum and its relation to quantum chaos is considered. 
A comparison with predictions from random matrix theory is presented. 
An analogous analysis is performed for the spectrum of the Gor'kov
representation of the fermionic action.
\end{abstract}
\maketitle
\sloppy

\section{Chiral and diquark condensate}

\begin{figure*}[pt]
  \label{fig1a}
  \hspace*{12mm}$\mu=0.2$\hspace*{20.3mm}$\mu=0.4$
  \hspace*{20.3mm}$\mu=0.6$\hspace*{20.3mm}$\mu=0.8$
  \hspace*{20.3mm}$\mu=1.5$\\[1mm]
  \centerline{\epsfig{figure=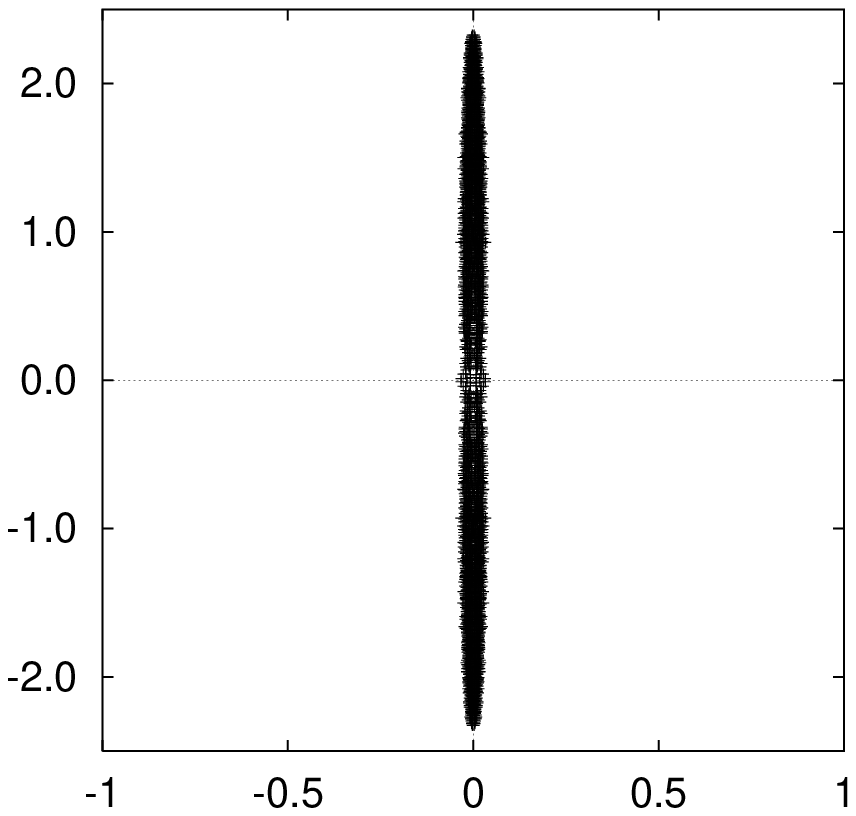,width=30mm}\hspace*{1mm}
    \epsfig{figure=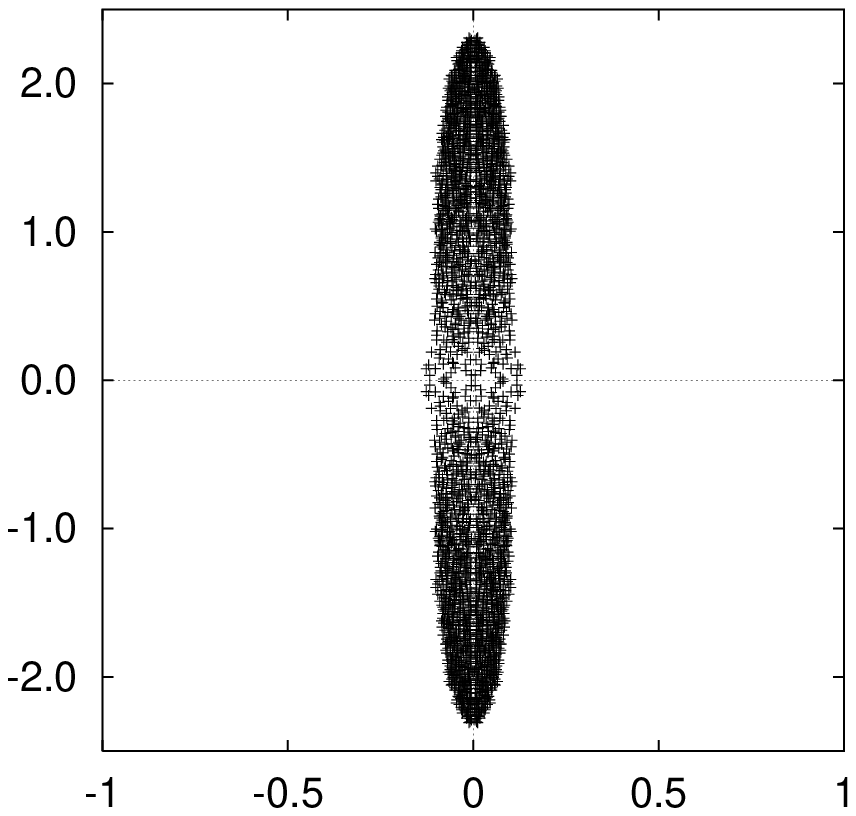,width=30mm}\hspace*{1mm}
    \epsfig{figure=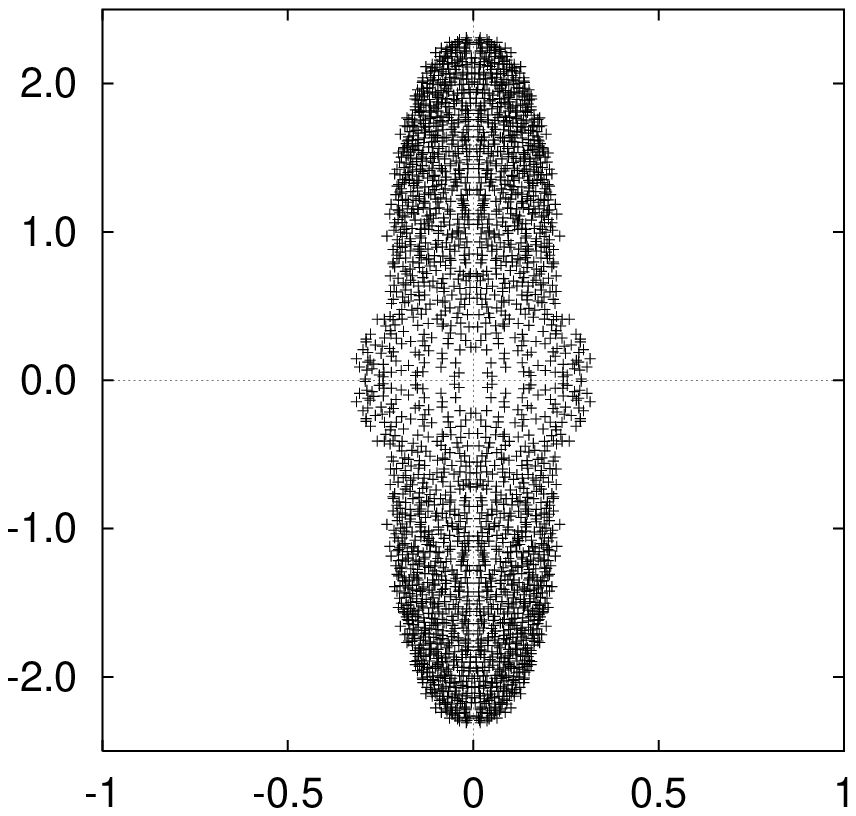,width=30mm}\hspace*{1mm}
    \epsfig{figure=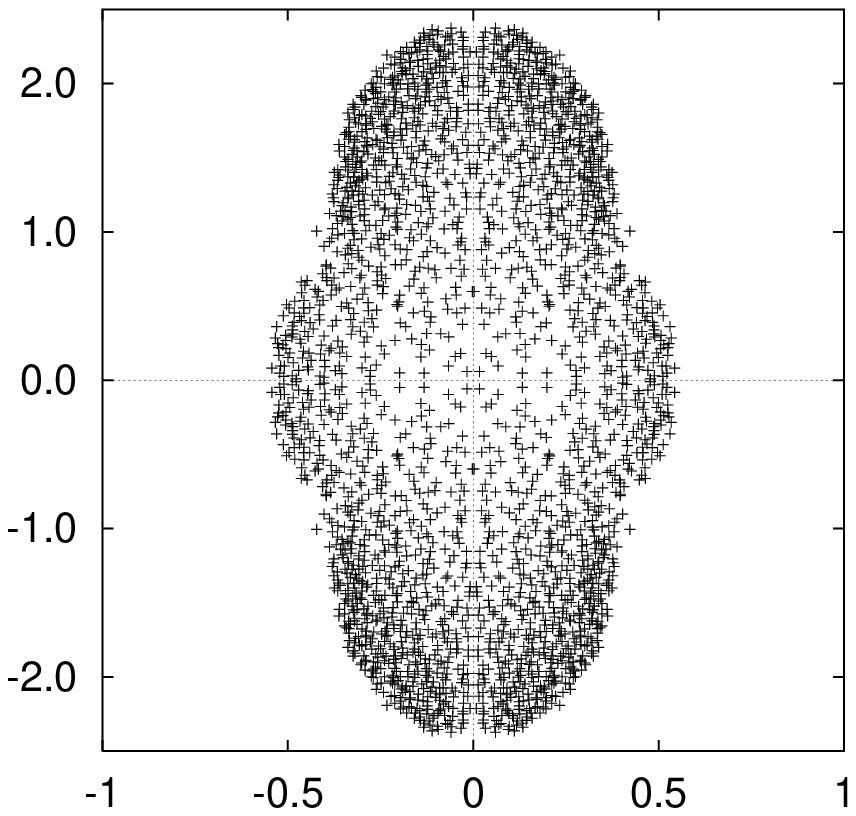,width=30mm}\hspace*{1mm}
    \epsfig{figure=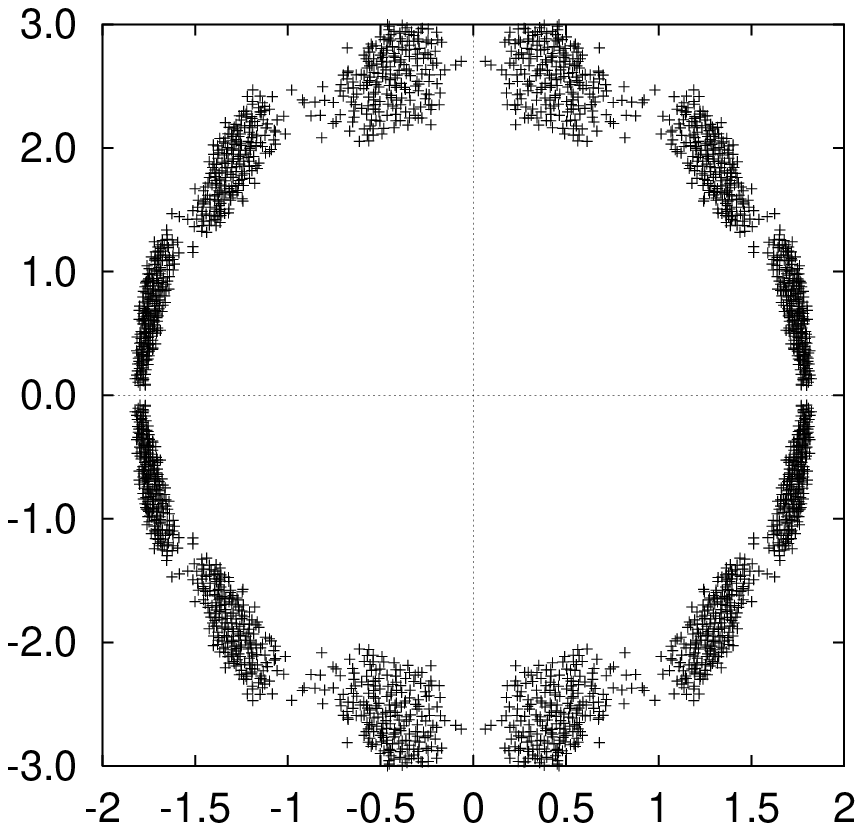,width=30mm}}\\[2mm]
  \centerline{\epsfig{figure=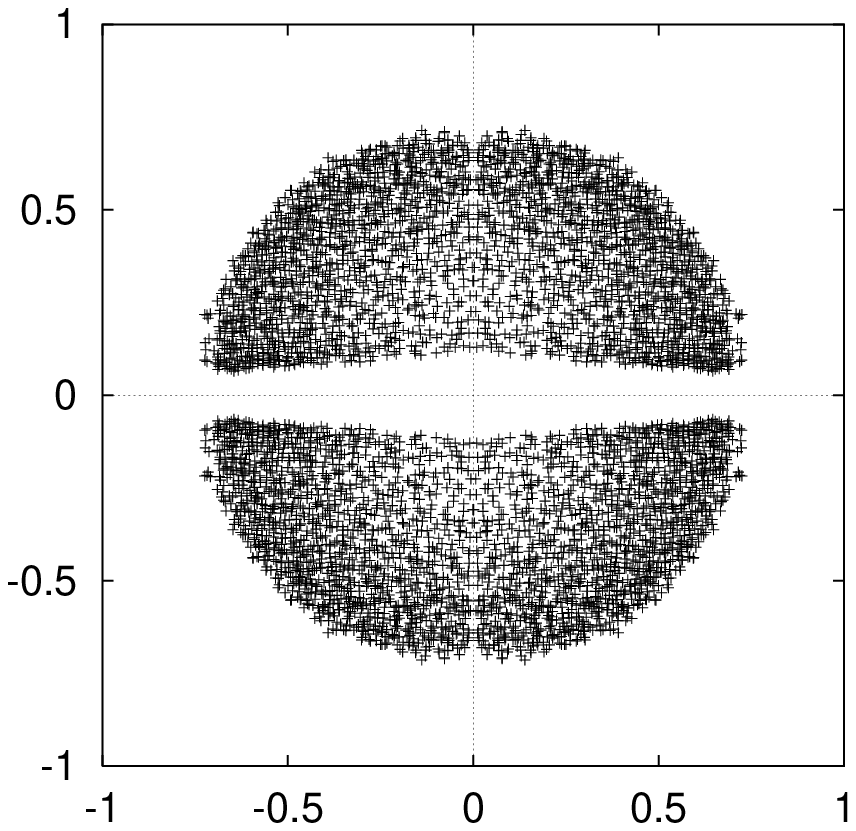,width=30mm}\hspace*{1mm}
    \epsfig{figure=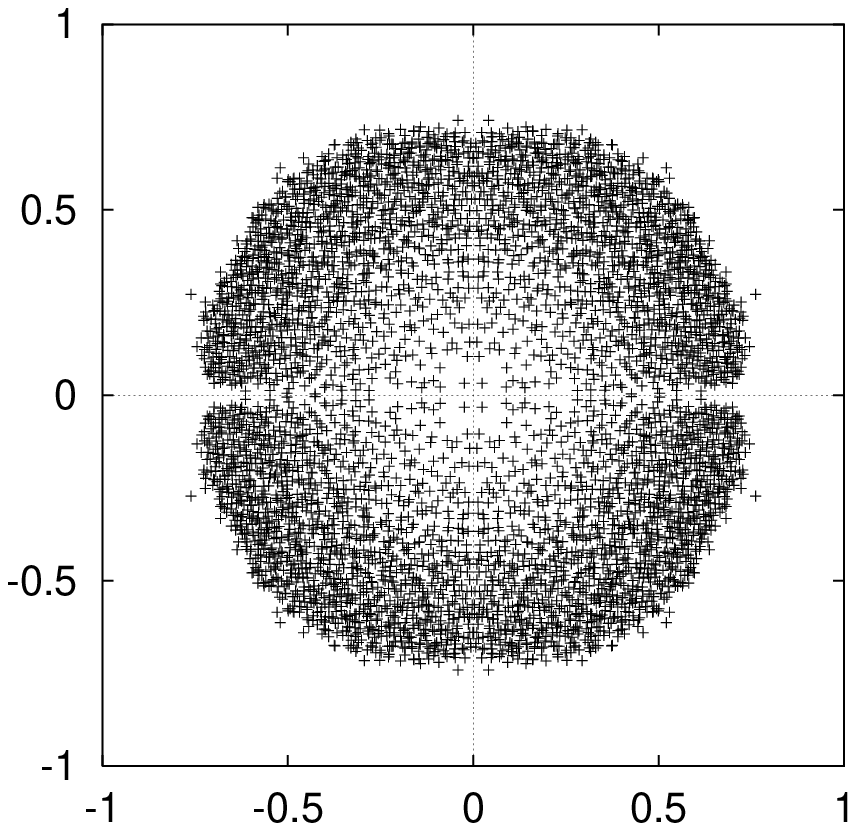,width=30mm}\hspace*{1mm}
    \epsfig{figure=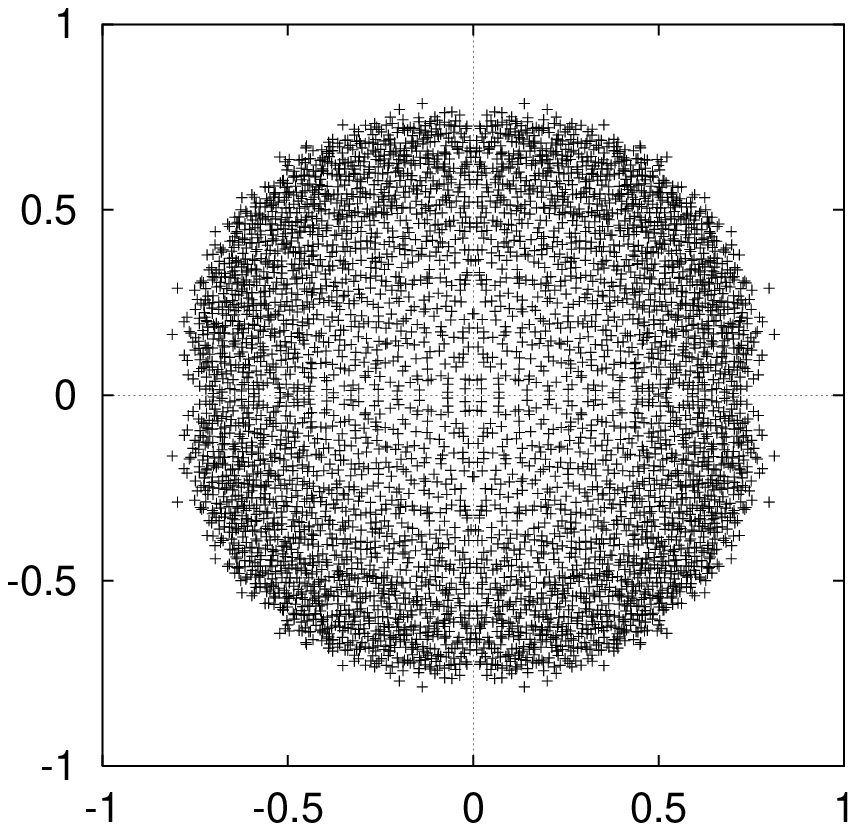,width=30mm}\hspace*{1mm}
    \epsfig{figure=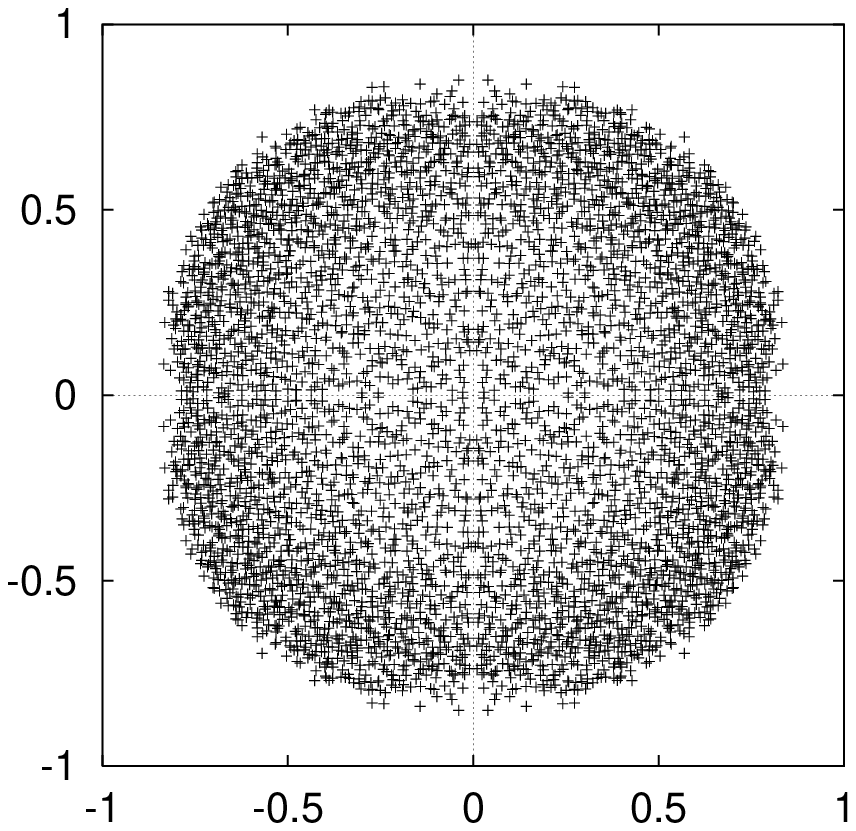,width=30mm}\hspace*{1mm}
    \epsfig{figure=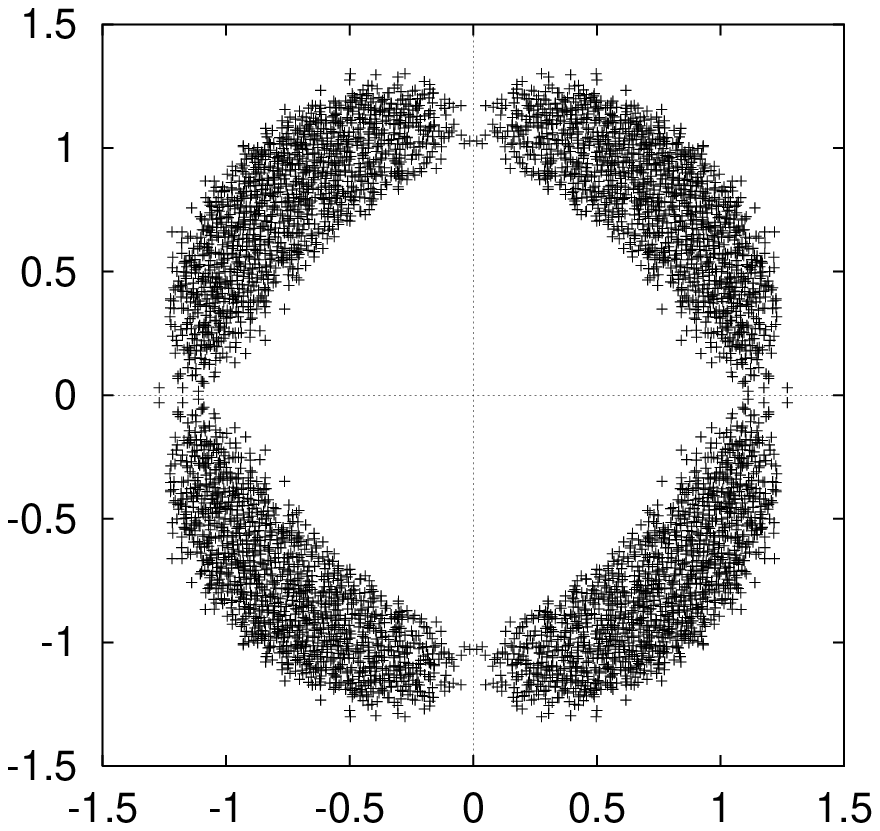,width=30mm}}
\vspace*{-10mm}
  \caption{Upper plots: Complex eigenvalues of the Dirac operator at
     various values of $\mu$ for a typical equilibrium configuration
     of two-color QCD (horizontal axes = real parts, vertical axes =
     imaginary parts, in units of $1/a$). Lower plots: Same for the
     ``Gor'kov operator''.}

\vspace*{8mm}

   \begin{center}
     \epsfig{figure=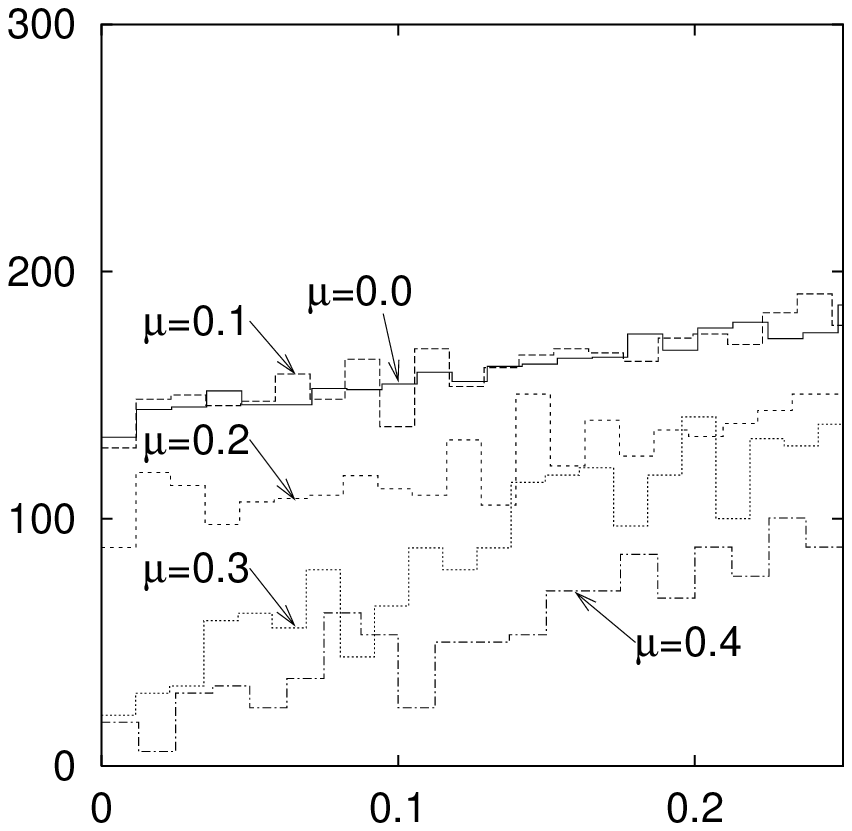,width=4.9cm,height=3.8cm}\hspace*{15mm}
     \epsfig{figure=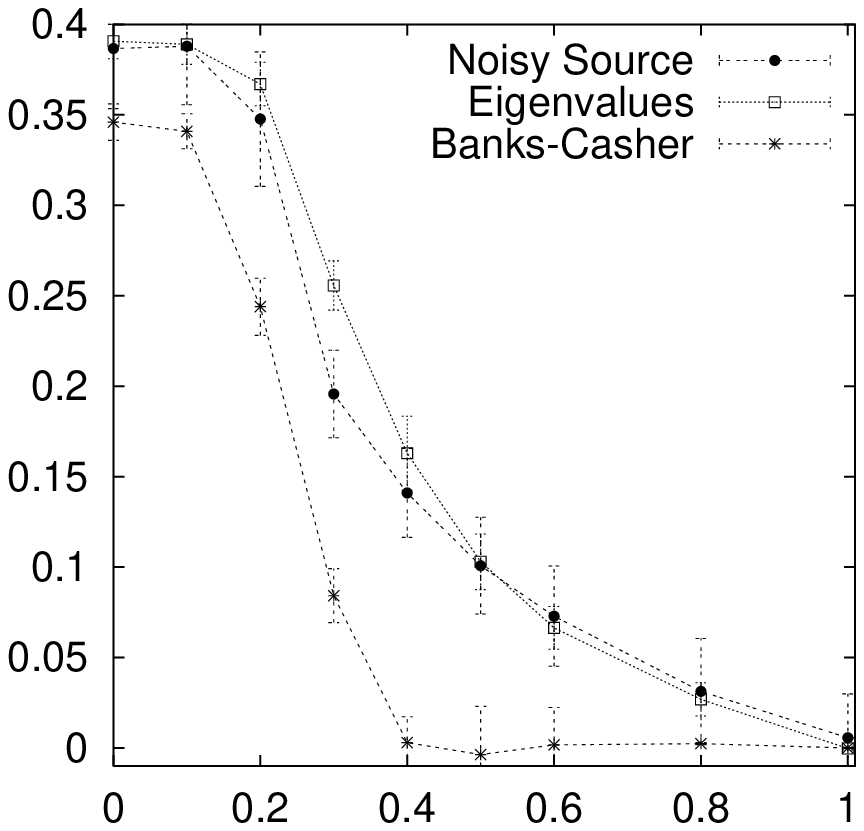,width=4.9cm,height=3.8cm}\\[-42mm]
     \hspace*{-50mm}\phantom{$\rho(y)$}\hspace*{50mm}\phantom{$\langle\bar\psi\psi\rangle$}\\[0.5mm]
     \hspace*{-54mm}$\rho(y)$\hspace*{59mm}$\langle\bar\psi\psi\rangle$\\[29mm]
     \hspace*{50mm}$y$\hspace*{68mm}$\mu$\\[2mm]
   \end{center}
   \begin{center}
     \epsfig{figure=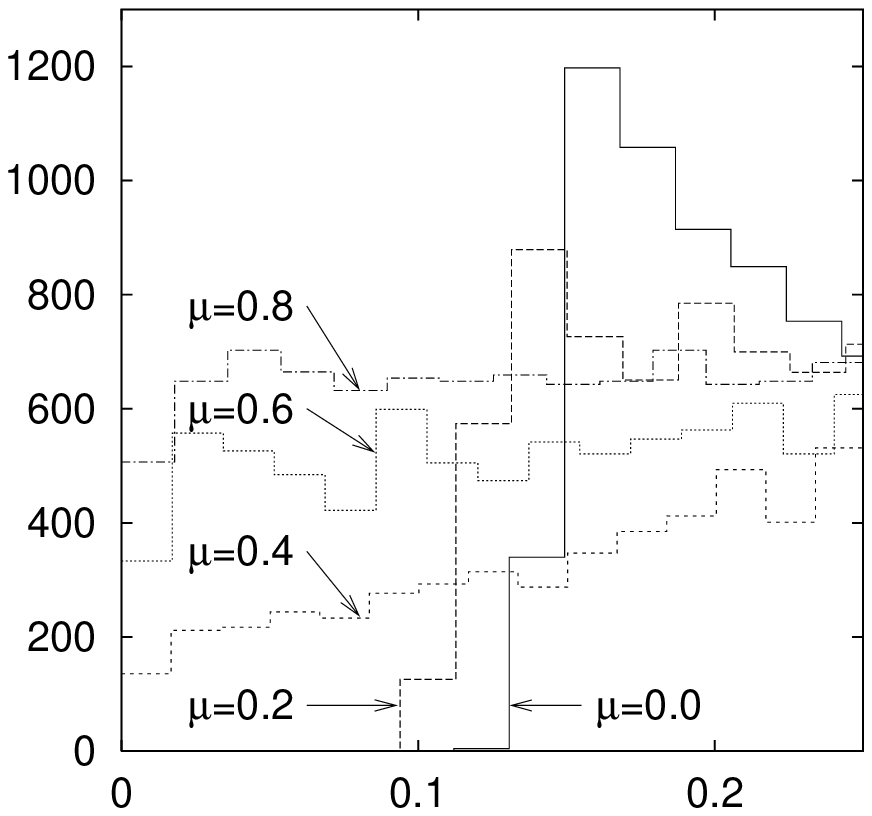,width=4.9cm,height=3.8cm}\hspace*{15mm}
     \epsfig{figure=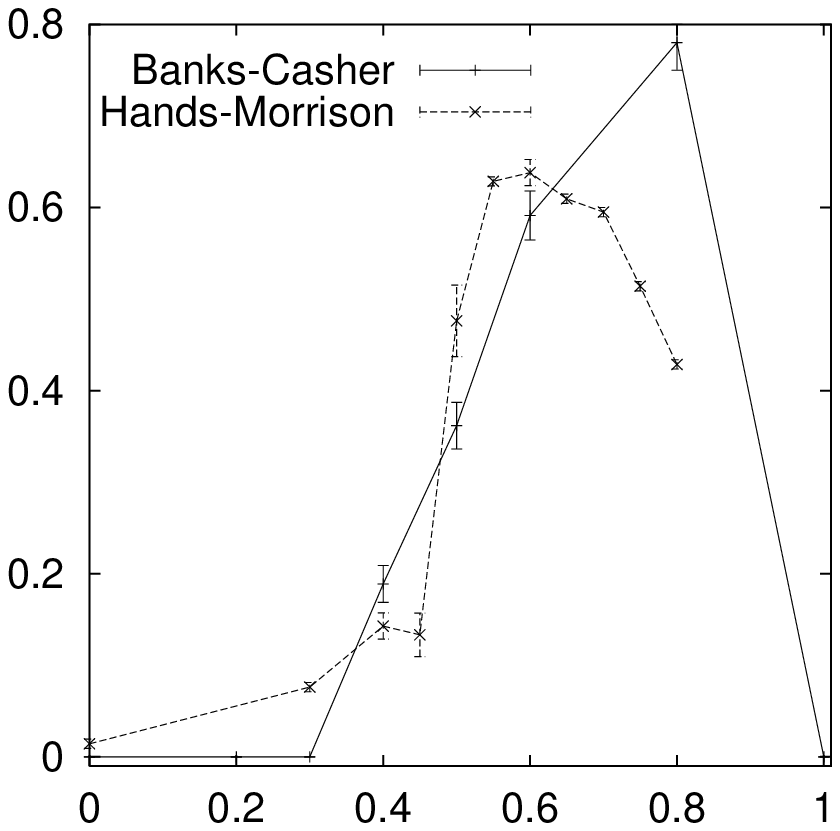,width=4.9cm,height=3.8cm}\\[-42mm]
     \hspace*{-50mm}\phantom{$\rho(y)$}\hspace*{50mm}\phantom{$\langle\psi\psi\rangle$}\\[0.5mm]
     \hspace*{-54mm}$\rho(y)$\hspace*{59mm}$\langle\psi\psi\rangle$\\[29mm]
     \hspace*{50mm}$y$\hspace*{68mm}$\mu$
   \end{center}
\vspace*{-10mm}
   \caption{Upper plots: Density $\rho(y)$ of small eigenvalues of the Dirac operator
        for two-color QCD on a $6^4$ lattice from $\mu=0$ to 0.4 (left).
        The loss of quasi-zero modes is accompanied by a vanishing
        of the chiral condensate. Chiral condensate extracted by different
        methods (right).
            Lower plots: Similar for the ``Gor'kov operator'' (see text).}
   \label{fig1}
\vspace*{-0.5mm}

\end{figure*}

The eigenvalues of the Dirac operator are of great
interest for important features of QCD. The
Banks-Casher formula \cite{Bank80} relates the Dirac eigenvalue density
$\rho(\lambda)$ at $\lambda=0$ to the chiral condensate,
$ \Sigma \equiv |\langle \bar{\psi} \psi \rangle| =
 \lim_{\varepsilon\to 0}\lim_{V\to\infty} \pi\rho (\varepsilon)/V$.
The microscopic spectral density, $ \rho_s (z) = \lim_{V\to\infty} 
 \rho \left( {z/V\Sigma } \right)/V\Sigma , $
should be given by the appropriate result of random matrix theory 
(RMT)~\cite{ShVe92}, which
also generates the Leutwyler-Smilga sum rules~\cite{LeSm92}.

A formulation of the QCD Dirac operator at chemical potential
$\mu\ne0$ on the lattice in the staggered scheme is
given by
\begin{eqnarray}
  \label{Dirac}
  M_{x,y}%(U,\mu)
\!\!\!&\!=\!&\!\!\!
  \frac{1}{2a}
\left\{ 
  \sum\limits_{\nu=\hat{x},\hat{y},\hat{z}}
  \left[U_{\nu}(x)\eta_{\nu}(x)\delta_{y,x\!+\!\nu}-{\rm h.c.}\right]
\right.
\nonumber\\
  &&\!\!\!\!\!\hspace*{-64pt}
\left.\phantom{\sum\limits_{\nu=\hat{x},\hat{y},\hat{z}}}
 +\!\!\left[U_{\hat{t}}(x)\eta_{\hat{t}}(x)e^{\mu}
    \delta_{y,x\!+\!\hat{t}}\!
    -\!U_{\hat{t}}^{\dagger}(y)\eta_{\hat{t}}(y)
    e^{-\mu}\delta_{y,x\!-\!\hat{t}}\right]
\!\right\}
\! ,
\end{eqnarray}
with the link variables $U$ and the staggered phases $\eta$.
The bilinear form of the fermionic action
$S_{ferm} = \bar\psi M \psi$ permits only to calculate the one-point
quark-antiquark function $\langle \bar\psi\psi \rangle$.
A possible extension of the Hilbert space is provided by the so-called
Gor'kov representation of the action
\begin{eqnarray}
\label{Gorkov}
S_{ferm}=(\bar\psi,\psi^{tr})\left(\matrix{\bar\jmath\tau_2&{1\over2}M\cr
-{1\over2}M^{tr}&j\tau_2\cr}\right)\left(\matrix{\bar\psi^{tr}\cr\psi\cr}
\right)\:,
\end{eqnarray}
with external source j, which allows to extract the diquark one-point
function  $\langle \psi\psi \rangle$~\cite{Morr99}.

The expression for the diquark condensate given in ~\cite{Morr99}
can be reformulated as
%\begin{equation}
$<\psi \psi> = \lim_{j \to 0} \lim_{V \to \infty} \langle 1/(2V)
\sum_n (\lambda_n^K + j)^{-1} \rangle$,
%\end{equation}
where the $\lambda_n^K$ are eigenvalues of the matrix
\begin{eqnarray}
\label{K}
K =\left(\matrix{\bar\jmath I &{1\over2}\tau_2 M \cr
-{1\over2} \tau_2 M^{tr} &j I \cr}\right)\:,
\end{eqnarray}
whose dependence on the diquark source $j$ is via $jI$.
The $\lambda_n^K$ then play, in the study of diquark condensation,
the same role as the eigenvalues of the Dirac operator in the
case of chiral condensation.
%{\em Mutatis mutandis}, all the interrelations of
%the Dirac eigenvalues and $\langle \bar \psi \psi \rangle$
%mentioned at the onset of this note should remain true for $K$ 
%and $\langle \psi \psi \rangle$. 
Finally we note that $K$ is  related to the ``Gor'kov operator''
in Eq.~(\ref{Gorkov}) by a multiplication with a constant matrix.
%: it is then reasonable to assume that the eigenvalues
%of $K$ and those of the ``Gor'kov operator'' describe the same critical behavior, and
In the following we will concentrate on the analysis of the Gor'kov
spectrum.

Our computations with gauge group SU(2) on a $6^4$ lattice
at $\beta=4/g^2=1.3$ have $N_f=2$ flavors
of staggered fermions of mass $m=0.07$. For this system the fermion
determinant is real and lattice simulations of the full theory with
chemical potential become feasible. They  exhibit a  phase
transition at $\mu_c \approx m_{\pi}/2 \approx 0.3$ where the chiral condensate 
(nearly) vanishes and a diquark condensate develops~\cite{Hand99}.

In the upper row of plots in Fig.~\ref{fig1a} we present the behavior
of the eigenvalues of the Dirac operator with increasing chemical
potential. A dilution of the eigenvalues around the origin is
observed. The lower row of plots depicts the eigenvalues of the ``Gor'kov
operator'' for $j=0$ as a function of $\mu$. One clearly sees that their density
at the origin increases except for large $\mu$.

In the upper left plot of Fig.~\ref{fig1} we compare the densities
of the small eigenvalues of the Dirac operator at $\mu=0$ to 0.4 on our
$6^4$ lattice, averaged over at least 160 configurations. Since the eigenvalues
move into the complex plane for $\mu > 0$, a band of width $\epsilon
= 0.015$ parallel to the imaginary axis is considered to construct
$\rho(y)$, i.e. $\rho(y)\equiv\int_{-\epsilon}^\epsilon
dx\,\rho(x,y)$, where $\rho(x,y)$ is the density of the complex
eigenvalues $x+iy$.

The density $\rho(y)$ is used to estimate a value for the chiral condensate
by  applying the Banks-Casher relation (which originally was derived
for real eigenvalues appearing in pairs of opposite sign). We further
employed the standard definition of the Green function~\cite{LeSm92} by
inverting the fermionic matrix with a noisy source and by computing its
eigenvalues exactly, respectively, to get
the condensate. Thus the chiral condensate for two-color QCD with
finite chemical potential was extracted by three methods. The
preliminary results for $<\bar \psi \psi>$ and its variance are shown
in the upper right plot of Fig.~\ref{fig1}.

In the lower left plot of Fig.~\ref{fig1} we show the densities of
the small eigenvalues of the ``Gor'kov operator'', averaged over
50 configurations. Again a band $\epsilon = 0.1$ is taken to 
construct $\rho(y)$ from the complex eigenvalues. This density is
employed in the spirit of the Banks-Casher relation to 
obtain a value for the diquark condensate. It compares well
with the results of Ref.~\cite{Morr99} from an inversion of the
$K$ matrix~(\ref{K}) with $m=0.05$ and $\beta=1.5$, see lower right plot
of Fig.~\ref{fig1}.

%Inserting the inverse
%eigenvalues into the formula of the Green's function~\cite{LeSm92}
%yields zero since the contributions cancel each other, and
%spontaneous symmetry breaking cannot be extracted
%from the finite system. Using the definition of
%the Green's function via external sources, a diquark condensate
%was computed in Ref.~\cite{Morr99}. These results (for $m=0.1$)
%are compared to the Banks-Casher formula in the

%%%%%%%%%%%%%%%%%%%%%%%%%%%%%%%%%%%%%%%%%%%%%%%%%%%%%%%%%%%%%%%%%%%%%%

\section{Quantum chaos}

The fluctuation properties of the eigenvalues in the bulk of the
spectrum have also attracted attention. It was shown in
Ref.~\cite{Hala95} for Hermitian Dirac operators that on
the scale of the mean level spacing they are described by RMT.
For example, the nearest-neighbor spacing
distribution $P(s)$, i.e. the distribution of spacings $s$ between
adjacent eigenvalues on the unfolded scale, agrees with the Wigner
surmise of RMT.  According to the Bohigas-Giannoni-Schmit
conjecture~\cite{Bohi84}, quantum systems whose classical counterparts are
chaotic have a nearest-neighbor spacing distribution given by RMT
whereas systems whose classical counterparts are integrable obey a
Poisson distribution, $P_{\rm P}(s)=e^{-s}$.  Therefore, the specific
form of $P(s)$ is often taken as a criterion for the presence or
absence of ``quantum chaos''.

For $\mu>0$, the Dirac operator
loses its Hermiticity properties so that its eigenvalues become
complex. We apply a
two-dimensional unfolding procedure~\cite{Mark99} to separate the
average eigenvalue 
density from the fluctuations and construct the nearest-neighbor
spacing distribution, $P(s)$, of adjacent eigenvalues in the complex
plane.
% Adjacent eigenvalues are defined to be the pairs for which the
%Euclidean distance in the complex plane is smallest.  
The data are then compared to analytical predictions of the Ginibre
ensemble~\cite{Gini65} of non-Hermitian RMT, which describes the
situation where the real and imaginary parts of the strongly
correlated eigenvalues have approximately the same average magnitude.
For this case $P(s)$ is given in Ref.~\cite{Grob88}.
%\begin{equation}
%  \label{Ginibre}
%  P_{\rm G}(s)=c \, p(cs)\:,
%\end{equation}
%with
%\begin{displaymath}
%  p(s)=2s\lim_{N\to\infty}\left[\prod_{n=1}^{N-1}e_n(s^2)\,e^{-s^2}\right]
%  \sum_{n=1}^{N-1}\frac{s^{2n}}{n!e_n(s^2)}\:,
%\end{displaymath}
%where $e_n(x)=\sum_{m=0}^n x^m/m!$ and $c=\int_0^\infty ds \, s \,
%p(s)=1.1429...$. 
For uncorrelated eigenvalues in the
complex plane, the Poisson distribution becomes
%\begin{equation}
%  \label{Poisson}
$  P_{\bar{\rm P}}(s)=(\pi/2)\,s\,\exp(-\pi s^2/4)$~\cite{Grob88}. 
%\end{equation}
This should not be confused with the Wigner distribution
for a Hermitian operator~\cite{Hala95}.

\begin{figure*}[t]
\begin{center}
\begin{tabular}{ccc}
  \hbox{\hspace{6.5mm} $\mu=0$}  &  \hbox{\hspace{2.75mm} $\mu=0.4$} &
  \hbox{\hspace{1.25mm} $\mu=3.0$ }\\[1.5mm]
  \epsfxsize=3.7cm\epsffile{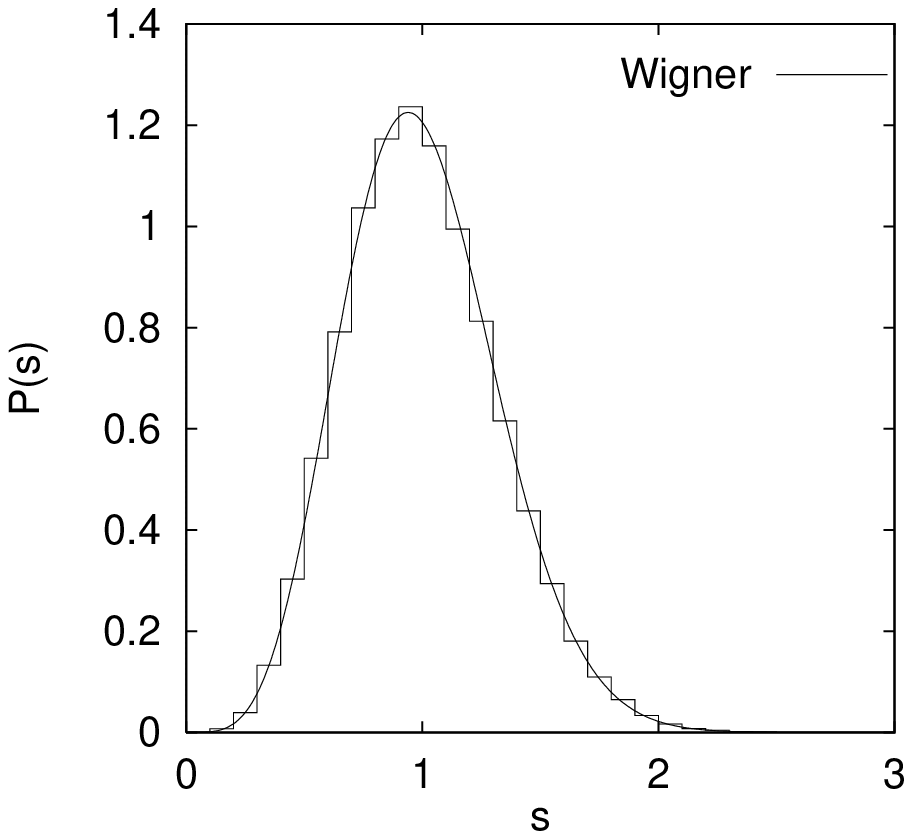} &
  \epsfxsize=3.7cm\epsffile{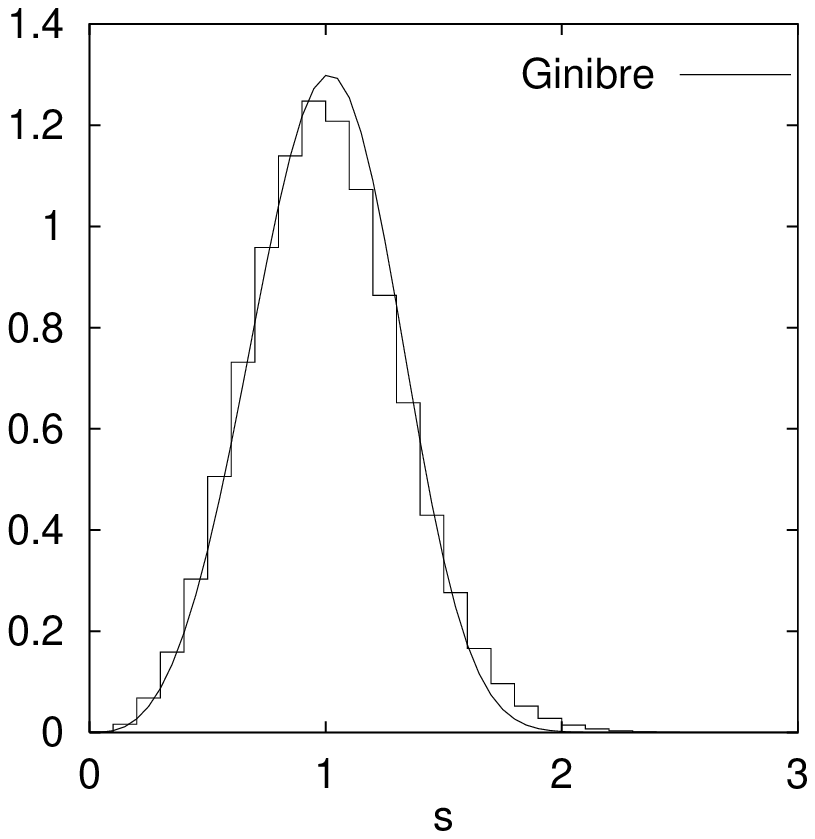} &
  \hbox{\hspace{-2.mm}\epsfxsize=3.7cm\epsffile{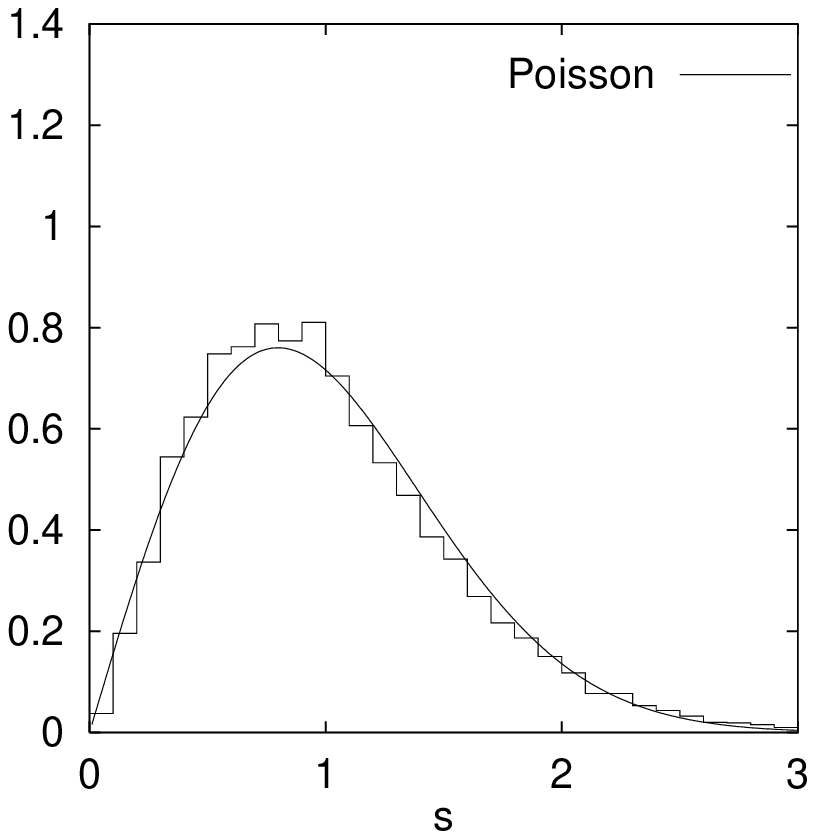}}\\
  \epsfxsize=3.7cm\epsffile{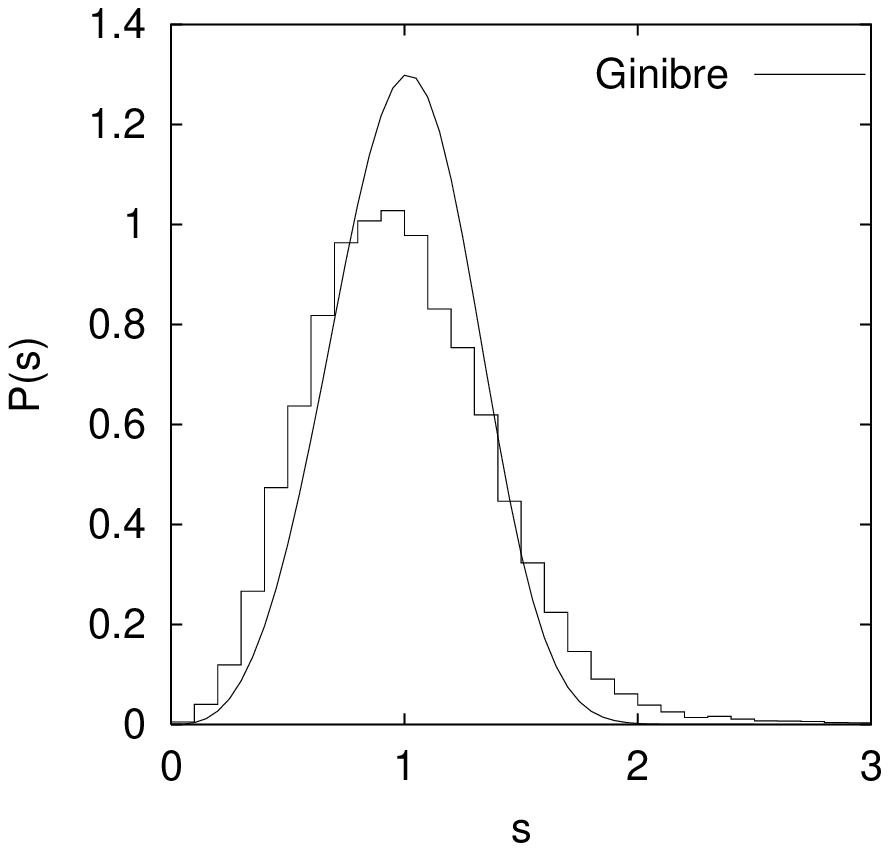} &
  \epsfxsize=3.7cm\epsffile{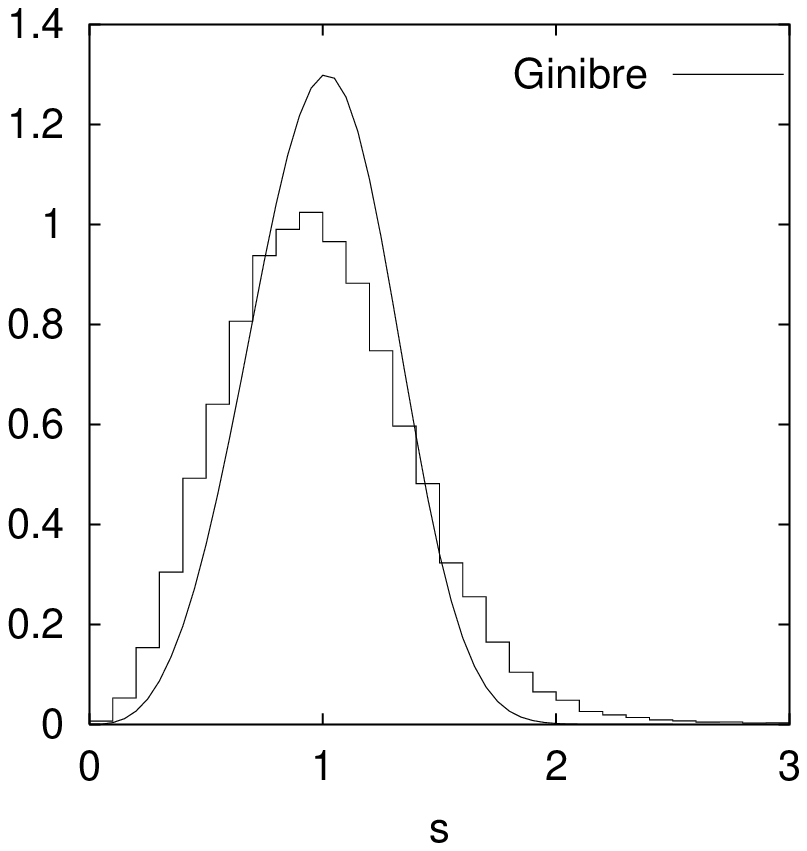} &
  \hbox{\hspace{-2.mm}\epsfxsize=3.7cm\epsffile{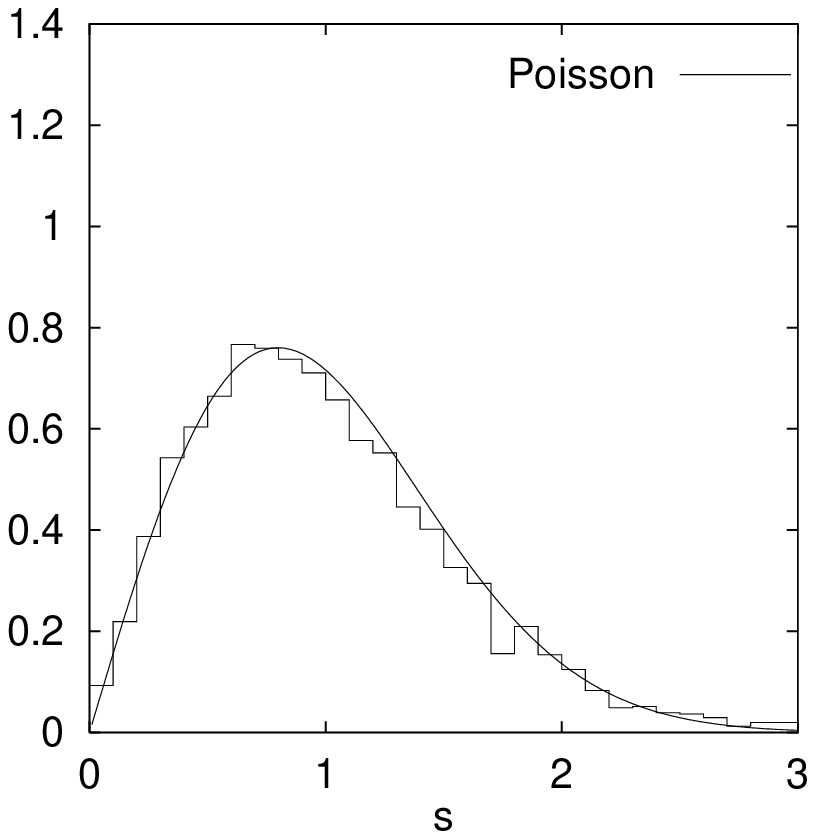}}
\end{tabular}
\end{center}
\vspace*{-14mm}
  \caption{Nearest-neighbor spacing distribution of the Dirac
    matrix (upper plots) and of the ``Gor'kov operator'' (lower plots)
    for two-color QCD with varying $\mu$.
    The curves are the Wigner distribution $P_{W} =
    262144/(729 \pi^3) s^4 \exp(-64/(9\pi)s^2)$,
    the Ginibre distribution of Ref.~\protect\cite{Grob88} and the
    Poisson distribution $  P_{\bar{\rm P}}(s)=(\pi/2)\,s\,\exp(-\pi s^2/4)$,
    respectively.}
  \label{fig3}
\vspace*{-4mm}
\end{figure*}

Our results for $P(s)$ of the Dirac operator are presented
in the upper row of Fig.~\ref{fig3}.  As a
function of $\mu$, we expect to find a transition from Wigner to
Ginibre behavior in $P(s)$~\cite{Mark99}. For the
symplectic ensemble of color-SU(2) with staggered fermions, the Wigner
and Ginibre distributions are very close to each other and thus hard
to distinguish. They are reproduced by our preliminary data for
$\mu=0$ and $\mu=0.4$, respectively. Even in the deconfined phase,
where the effect of the chemical potential might order the system, 
the gauge fields  retain a considerable degree of randomness, which
apparently gives rise to quantum chaos.

For $\mu > 1.0$, the lattice results for $P(s)$ deviate substantially
from the Ginibre distribution and could be interpreted as Poisson
behavior, corresponding to uncorrelated eigenvalues. A plausible
explanation of the transition to Poisson behavior is provided by the
following two observations. First, for large $\mu$ the terms
containing $e^\mu$ in Eq.~(\ref{Dirac}) dominate the Dirac matrix
giving
rise to uncorrelated eigenvalues. Second, for large $\mu$ the fermion density
on the finite lattice reaches saturation due to the limited box size and
the Pauli exclusion principle.

The results for $P(s)$ of the ``Gor'kov operator'' are plotted in 
the lower row of Fig.~\ref{fig3}. Since the eigenvalues are complex
also for $\mu=0$, we find approximately Ginibre behavior corresponding to
non-Hermitian RMT. For $\mu = 3.0$, $P(s)$ takes a Poisson distribution
associated with uncorrelated complex eigenvalues.

{\it Acknowledgments:}
This study was supported in part by FWF project P14435-TPH.
We thank the ECT*, Trento, for hospitality during
various stages of this work. In addition M.-P.L. thanks the 
Institute for Nuclear Theory at the University of Washington
and the Department of Energy for partial support. 
Finally, we wish to thank B.A. Berg, P.H. Damgaard, M.A. Halasz,
S. Hands, T. Sch\"afer,  D.K. Sinclair,  M.A. Stephanov, J.J.M. Verbaarschot
and  T. Wettig for discussions~\cite{osaka}.

%%%%%%%%%%%%%%%%%%%%%%%%%%%%%%%%%%%%%%%%%%%%%%%%%%%%%%%%%%%%%%%%%%%%%%%%%%%%%%%%

\end{document}